%% file: main.tex
\documentclass[10pt,twocolumn,superscriptaddress, aps, prl]{revtex4-1}

\usepackage{natbib}
\usepackage{graphicx}
\usepackage{amsmath}
\usepackage{fullpage}
\usepackage{hyperref}
\usepackage{bm}

\newcommand{\eps}{\varepsilon}

\hypersetup{
    colorlinks=true,       % false: boxed links; true: colored links
    linkcolor=black, %blue,          % color of internal links
    citecolor=blue, %black,        % color of links to bibliography
    filecolor=magenta,      % color of file links
    urlcolor=cyan           % color of external links
}

\begin{document}

\title{Vertically-coupled dipolar exciton molecules} 

\author{Kobi Cohen}
\affiliation{Racah Institute of Physics, The Hebrew University of Jerusalem, Jerusalem 9190401, Israel}
\author{Maxim Khodas}
\affiliation{Racah Institute of Physics, The Hebrew University of Jerusalem, Jerusalem 9190401, Israel}
\author{Boris Laikhtman}
\affiliation{Racah Institute of Physics, The Hebrew University of Jerusalem, Jerusalem 9190401, Israel}
\author{Paulo V. Santos}
\affiliation{Paul-Drude-Institut f\"{u}r Festk\"{o}rperelektronik, Hausvogteiplatz 5-7, 10117 Berlin, Germany}
\author{Ronen Rapaport}
\affiliation{Racah Institute of Physics, The Hebrew University of Jerusalem, Jerusalem 9190401, Israel}

\begin{abstract}

While the interaction potential between two dipoles residing in a single plane is repulsive, in a system of two vertically adjacent layers of dipoles it changes from repulsive interaction in the long range to attractive interaction in the short range. Here we show that for dipolar excitons in semiconductor heterostructures, such a potential may give rise to bound states if two such excitons are excited in two separate layers, leading to the formation of vertically coupled dipolar exciton molecules. Our calculations prove the existence of such bound states and predict their binding energy as a function of the layers separation as well as their thermal distributions. We show that these molecules should be observed in realistic systems such as semiconductor coupled quantum well structures and the more recent van-der-Waals bound heterostructures. Formation of such molecules can lead to new effects such as a collective dipolar drag between layers and new forms of multi-particle correlations, as well as to the study of dipolar molecular dynamics in a controlled system.

\end{abstract}

\maketitle

Two-dimensional (2D) fluids of quantum dipoles are currently central 
to recent theoretical and experimental efforts to explore many-body physics in the quantum regime. Research on such fluids ranges from atomic and molecular systems \cite{lahaye_physics_2009, jin_polar_2011} to condensed-matter semiconducting heterostructure systems \cite{eisenstein_bose-einstein_2004, butov_formation_2004, rapaport_charge_2004, cristofolini_coupling_2012,kazimierczuk_giant_2014}.
In the latter, dipolar excitons, also known as indirect excitons (IXs), are quasi-particles composed of a Coulomb-bound electron-hole pair, where each of the constituents resides in a separate material layer. Excitation of many such IXs results in the formation of a two-dimensional (2D) layer of oriented dipoles, with a repulsive interaction between particles. 
%which scales as $\sim d^2/r^3$ ($d$ is the dipole size and $r$ the in-plane distance between dipoles).
Due to their bosonic character and relatively long range interaction, these particles display a rich phase diagram, with collective phases and correlation regimes, such as liquidity, dark condensation, and complex spin textures \cite{combescot_bose-einstein_2007, laikhtman_exciton_2009, shilo_particle_2013, stern_exciton_2014, cohen_spontaneous_2015, high_spontaneous_2012, high_spin_2013} which are absent in weakly interacting Bose gases.  
In the last decade, there has been a significant progress in the formation, control, and understanding of dipolar IX quantum fluids in GaAs-based bilayers, consisting of biased double quantum well (DQW) structures. Recently, new material systems such as $GaN$-based polar quantum wells \cite{fedichkin_transport_2015} and van-der-Waals (vdW) heterostuctures \cite{mak_atomically_2010, rivera_observation_2015} enable formation of dipolar IXs at elevated temperatures, with dipole sizes ranging from few to tens of nano-meters. With such systems, the collective quantum phenomena might even be observed up to room temperature \cite{fogler_high-temperature_2014}.
In all these systems the dominant in-plane repulsion limits the formation of more complex structures, such as bi-excitons or charged complexes, to a narrow, hard-to-access parameters range and to very low binding energies \cite{yudson_charged_1996,lozovik_superfluidity_1999, schindler_analysis_2008, lee_exciton-exciton_2009}, suppressing potential observation of interesting phases that may arise from such pair correlations.

\begin{figure}[ht]
\includegraphics[width=0.4\textwidth]{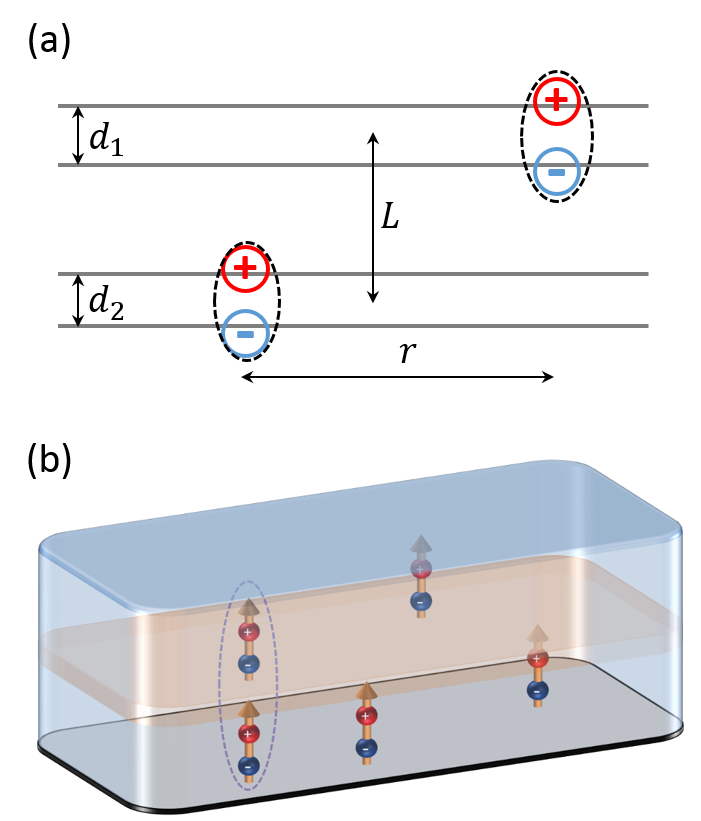}
\caption{\label{fig:dd} (a) A system of two dipoles of size $d_1, d_2$ residing in different layers. $L$ is the separation between the centers of the dipoles and $r$ is the in-plane distance. (b) An illustration of a vertical IXs molecule (dashed oval) together with unbound IXs in the two separated layers.}
\end{figure}

Here, we propose a \textit{double} bilayer structure, composed of two vertically stacked DQWs separated by a small vertical distance $L$ (see Fig.~\ref{fig:dd}(a)). This proximity results in an \textit{attractive} part for the \textit{out of plane} dipole-dipole interaction between the dipolar IXs formed in each of the bilayers.
As we show hereafter, this attractive interaction results in the formation of vertical IXs molecules, which are composed of bound pairs of two IXs from the two vertically coupled DQWs as illustrated in Fig.~\ref{fig:dd}(b).
This finding opens new possibilities, such as observing a transition from unbound, single IXs gas to a molecular gas when the temperature goes down or as the density is modified, and the possibility of attractive dipolar drag between IXs in the two coupled layers.

In the case of two oriented IXs residing in a single layer, there is a repulsive force between them at all lateral separation distances $r$. This repulsion gives rise to the well-known blue-shift of the exciton emission spectrum \cite{laikhtman_exciton_2009}.
However, the electrostatic interaction between two oriented IXs residing in \textit{different} yet proximate layers is qualitatively different: for small lateral distance $r$ the two IXs experience electrostatic attraction, that gradually changes to repulsion for large lateral distances.
This Coulomb interaction potential $U_{d}(r)$ between the two IXs is given by
\begin{align}
U_d(r) = &\frac{e^{2}}{\varepsilon}
    \nonumber    \Bigg[ \frac{1}{\sqrt{[L + (d_{1} - d_{2})/2]^{2} + r^{2}}} \\
	\nonumber	& +\frac{1}{\sqrt{[L - (d_{1} - d_{2})/2]^{2} + r^{2}}} \\
	\nonumber	& -\frac{1}{\sqrt{[L + (d_{1} + d_{2})/2]^{2} + r^{2}}} \\
				& -\frac{1}{\sqrt{[L - (d_{1} + d_{2})/2]^{2} + r^{2}}} \Bigg],
\label{eq:1}
\end{align}
where $d_{1,2}$ are the vertical sizes of each of the two IXs, $L$ is the vertical separation between the centers of the dipoles and $\varepsilon$ is the dielectric constant 
(it is implied here that $L>(d_{1}+d_{2})/2$).

\begin{figure}[ht]
\includegraphics[width=0.40\textwidth]{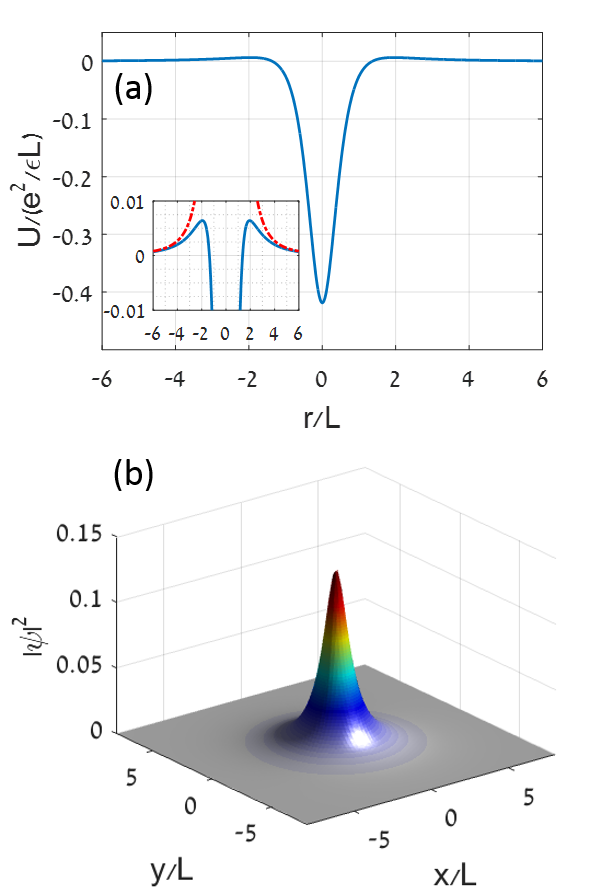}
\caption{\label{fig:u-dd} (a) Interaction energy of the two dipoles shown in Fig.~\ref{fig:dd}, in units of $e^{2}/\varepsilon L$, as a function of the lateral distance measured in units of the vertical separation between the two quantum double wells, $r/L$. Here the sizes of the dipoles are $d_{1}/L=4/9$, $d_{2}/L=7/18$.
The inset shows a zoom around zero energy to emphasize the transition from attraction to repulsion beyond $r\sim L$, together with a comparison to a pure dipole-dipole repulsive interaction $\sim d_1d_2/r^3$ (red, dash-dot).
(b) An exemplary numerical solution of the Schr\"odinger equation showing the probability density of the relative in-plane displacement $\vec{\rho}=(x/L,y/L)$ of the two excitons in the bound state, with the potential given in (a), assuming a GaAs material system with $L=300\AA$.}
\end{figure}

The potential in Eq.~\eqref{eq:1} is plotted in Fig.~\ref{fig:u-dd}(a) and has the following generic features: it reaches a minimum at zero lateral separation where $U_d(0)<0$, and remains negative up to a distance of the order of $L$, signifying attraction for $r \lesssim L$.
As for larger distances where $r \gtrsim L$, the potential $U_d(r)$ turns repulsive and decays as $U_d(r) \approx d_1 d_2 / r^3$ as expected for the interaction of the two in-plane aligned dipoles.
Finally, we note that the potential in Eq.~\eqref{eq:1} is neutral on average,
$\int d^2 r U_d(r) = 0$.

Any net attraction in two dimensions results in a bound state \cite{[{Binding energies for the 1D and 2D cases are calculated in }] [{. See problems after Sec.45, pp.162 and 163.}] landau_qm}. However, the neutral potential $U_d(r)$ is also known to produce a weakly bound state \cite{Simon1976, klawunn_two-dimensional_2010}. 
%In this case, the binding energy can be exponentially small and is estimated by $E_b \propto \exp\left[ - C(m/\hbar^2)^2 |U_d(0)|^{-2} L^{-4} \right]$, where $L$ is the potential range, in our case the inter-bilayer separation, and $C$ is a constant of order one (we note that the quoted estimate of the binding energy is good only for a very weak interaction potential). 
The question is therefore can experimentally significant bound states of this type exist? We thereafter consider two realistic systems of vertically coupled indirect excitons: electrically biased GaAs coupled quantum wells and vdW heterostructures. As we demonstrate below, the binding energies of these vertically coupled IXs molecules, which we denote by $\mathrm{IX}_M$, are significant and experimentally relevant. We further argue that the formation of complexes which consist of more than two IXs is impossible in the considered double bilayer system.

The Schr\"odinger equation for the relative in-plane center-of-mass separation between the two IXs, $\vec{\rho} = \vec{r}/L$, reads
\begin{equation}\label{eq:schrodinger}
	-\frac{\hbar^2}{2\mu}\frac{1}{L^2}\nabla^2\psi(\vec{\rho})+\frac{e^2}{\varepsilon L}\tilde{U_d}(\vec{\rho})\psi(\vec{\rho})=E\psi(\vec{\rho})
\end{equation}
where $\mu$ is the reduced mass of the two IXs and $\tilde{U_d}$ is the potential given in Eq. \eqref{eq:1} with $e^2/\varepsilon L$ factored out.
First we estimate the typical scales of the problem.
For GaAs heterostructures $L\sim 30$nm, while for vdW material systems the typical inter-layer separation is $L \sim 1$nm \cite{Splendiani2010}.
The Coulomb interaction is characterized by the energy scale $e^{2}/\varepsilon L \approx 12\mathrm{meV}/(L/10\mathrm{nm})$ for GaAs-based bilayers (using $\varepsilon\approx 12$), and by $e^{2}/\varepsilon' L \approx 230 \mathrm{meV}/(L/1\mathrm{nm})$ in vdW coupled heterostructures (taking the effective dielectric constant $\varepsilon'\approx (1+ \epsilon_{Si})/2 \approx 6.35$ for a system placed on a Si substrate with $\epsilon_{Si} =11.7$ \cite{rivera_observation_2015}). 
The kinetic energy scale in GaAs-based bilayers is $\hbar^2/ \mu L^2  \approx 8 \mathrm{meV}/(L/10\mathrm{nm})^2$, where the reduced mass is $\mu = \frac{1}{2}(m_e + m_{hh})$ (with the electron mass $m_e = 0.067 m_0$ and the heavy-hole mass $m_{hh}= 0.11 m_0$, and $m_0$ is the bare electron mass). For vdW heterostructures the effective exciton mass is approximately $m_0$ \cite{peelaers_effects_2012}, so the kinetic energy scale is estimated by $\hbar^2/\frac{1}{2}m_0 L^2 \approx 150 \mathrm{meV}/(L/1\mathrm{nm})^2$.

\begin{figure}
\includegraphics[width=0.4\textwidth]{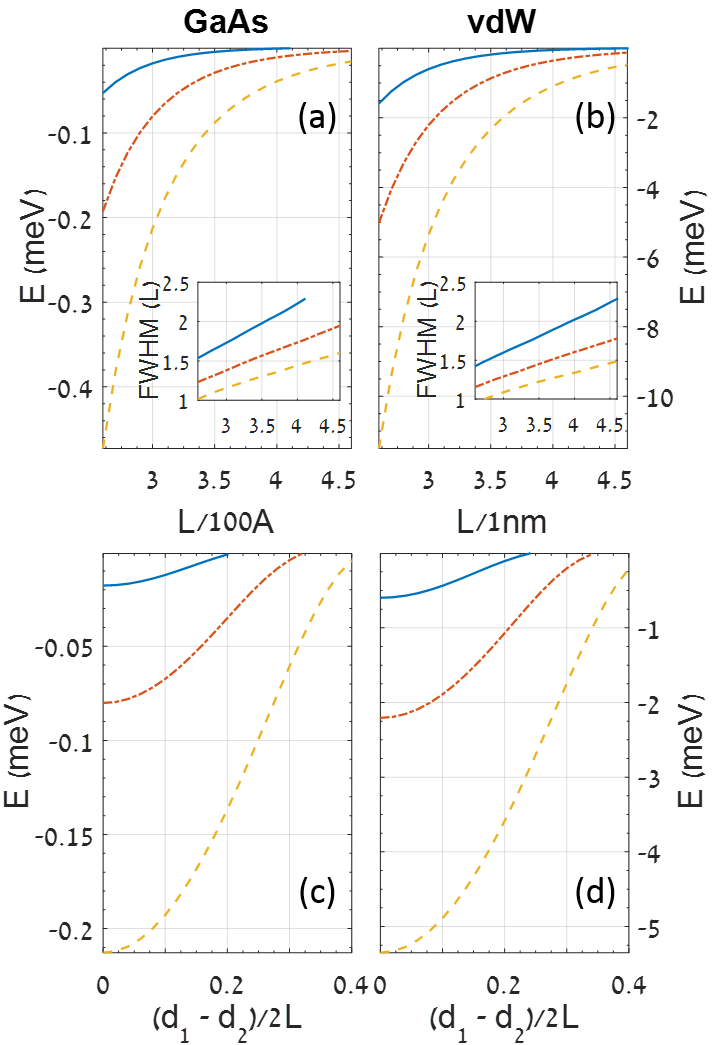}
% \caption{\label{fig:binding_symm}
\caption{\label{fig:binding} 
(a,b) Binding energy of IX$_M$ molecule as a function of the dipoles separation $L$ in GaAs and vdW materials. (a) $d_1=d_2=120 \AA$ (blue, solid), $d_1=d_2=140 \AA$ (red, dash-dot) and $d_1=d_2=160 \AA$ (yellow, dashed). (b) $d_1=d_2=1.2$nm (blue, solid), $d_1=d_2=1.4$nm (red, dash-dot) and $d_1=d_2=1.6$nm (yellow, dashed). Insets: full width at half maximum (FWHM) of the probability density $|\psi(\vec{\rho})|^2$, measured in units of $L$.
(c,d) Binding energy of IX$_M$ molecule formed in GaAs and vdW materials as a function of the structure asymmetry, $(d_1 - d_2)/2L$ at fixed $L = 300 \AA$ (c) and $L = 3$nm (d).
In both panels $(d_1+d_2)/2 L$ was fixed to $6/15$ (blue, solid), $7/15$ (red, dash-dot) and $8/15$ (yellow, dashed).
%The binding energy is maximal for symmetric structures, and the bound state is practically lost for some critical degree of asymmetry.
}
\end{figure}

We find the bound state of the two IXs by numerically solving Eq. \eqref{eq:schrodinger}. An exemplary solution of the probability density $|\psi(\vec{\rho})|^2$ in the lowest energy bound state is shown in Fig.~\ref{fig:u-dd}(b). Below we discuss the dependence of the binding energy of IX$_M$ on a range of (realistic) geometrical parameters for the two studied exemplary material systems (GaAs and vdW).
%We begin with the dependence of the binding energy on the separation $L$ between the dipoles.
For the symmetric configuration $d_1 = d_2$, the calculated binding energy as a function of the dipoles separation $L$ is shown in Fig.~\ref{fig:binding}(a,b) for three different dipole sizes. The binding energy increases (in absolute value) with decreasing inter-layers separation $L$ and larger dipole size $d$, as the attraction between two aligned dipoles scales like $d^2/L^3$.
In addition, the insets of Fig.~\ref{fig:binding}(a,b) show the full width at half maximum (FWHM) of the probability density, which is a measure for the molecules' Bohr radius. One can see that the FWHM is of the same order of $L$ and increases with it. The calculation is performed in a limited range of inter-bilayer separation $L$, since for $L$ values that are too small %(and yet within the allowed range for $L$) 
the binding energy tends to exceed that of the unbound IX itself, thus making our treatment inaccurate. Furthermore, from the practical side, making $L$ too small might increase the electron inter-bilayer tunneling rate.

% \begin{figure}[ht]
% \includegraphics[width=0.45\textwidth]{fig4_binding_d1d2.png}
% \caption{\label{fig:binding_asymm}
% (Color online) The binding energy of IX$_M$ molecule formed in GaAs (left panel) and vdW materials (right panel) as a function of the structure asymmetry, $(d_1 - d_2)/2L$ at fixed $L = 300 \AA$ (left) and $L = 3$nm (right).
% In both panels $(d_1+d_2)/2 L$ was fixed to $6/15$ (blue, solid), $7/15$ (red, dash-dot) and $8/15$ (yellow, dashed). The binding energy is maximal for symmetric structures, and the bound state is practically lost for some critical degree of asymmetry.}
% \end{figure}

Next we study the effect of asymmetry between the two bilayers.  Asymmetric layers are of practical interest and importance since different layer thicknesses leads to different exciton energies so that the excitons in the two layers can be optically addressed and resolved independently. Fig.~\ref{fig:binding}(c,d) shows the calculated binding energy as a function of the difference $d_1-d_2$, for a few fixed values of $d_1+d_2$ and $L$. The dependence of binding energy on $d_1 \pm d_2$ at fixed $L$ is qualitatively understood as follows:
the overall increase in binding energy with $d_1+d_2$ is explained by the growing attraction between the electron and the hole in the upper and the lower bilayer respectively, given explicitly by the last term of Eq.~\eqref{eq:1}.
%(referring to a configuration as in Fig.~\ref{fig:dd}). Such attraction is given explicitly by the last term of Eq.~\eqref{eq:1}.
%At the same time, the vertical distances between the carriers of the same kind for a given $d_1 - d_2$ are fixed. Therefore, the repulsion part of the interaction contained in the first two terms of Eq.~\eqref{eq:1} is unaffected by changing $d_1 + d_2$.
The decrease in binding energy with $d_1-d_2$ at fixed $d_1+d_2$ and $L$ is because the attraction part above remains unaffected by $d_1-d_2$, while the repulsion part contained in the first two terms of Eq.~\eqref{eq:1} grows.
%Indeed, the distance between the nearest charges of the same kind decreases.
%For $d_1>d_2$ ($d_2>d_1$) it is the distance between the electrons (holes) that shrinks. The dominant repulsion interaction is then described by the second (first) term of the potential, Eq.~\eqref{eq:1}. 
It follows that symmetric structures yields the maximum binding and are in principle favorable, however practical considerations might require a slightly asymmetric structure.

\begin{figure}
\includegraphics[width=0.4\textwidth]{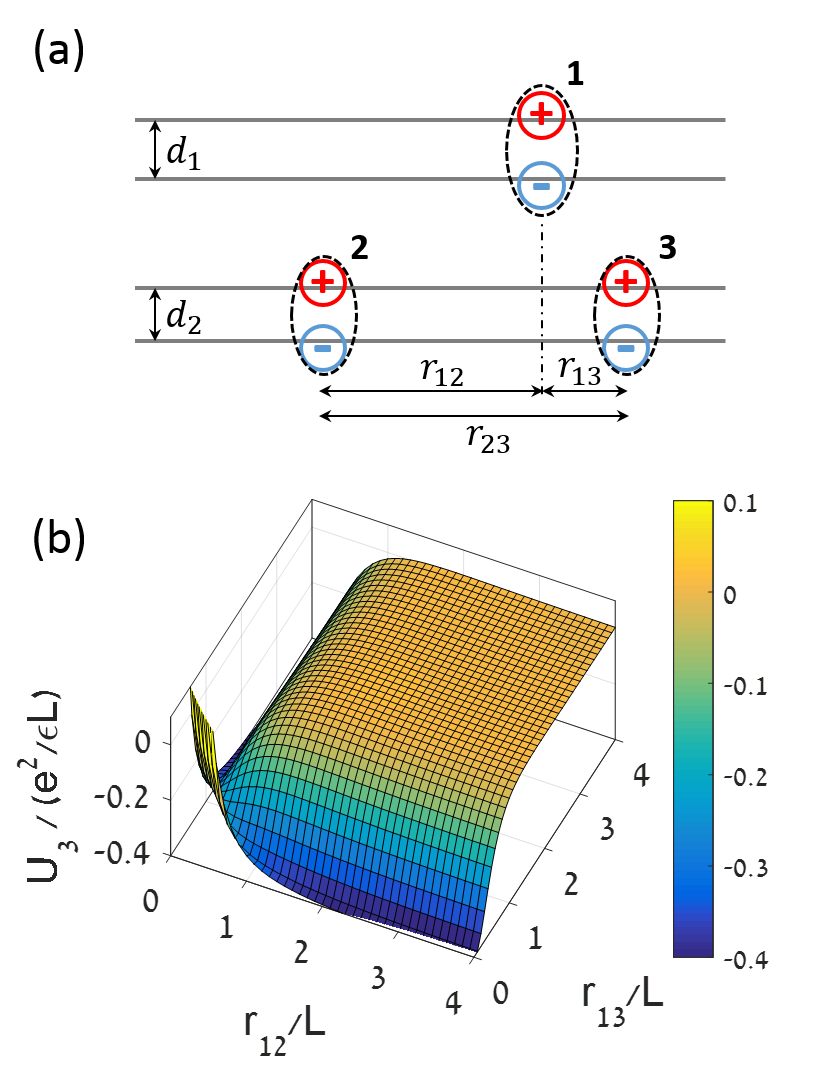}
\caption{(a) A system of three IX's, one resides in the upper DQW and the two others are in the lower DQW, with lateral distances of $r_{12},r_{13},r_{23}$ respectively. The optimal configuration minimizing the repulsion for given values of $r_{12}$ and $r_{13}$ requires that the three excitons reside on the same straight line (when viewed from the top), such that $r_{23} = r_{12} + r_{13}$.
(b) The interaction potential of three dipolar excitons $U_{3}(r_{12},r_{13},r_{23}=r_{12}+r_{13})$.
The minimum is reached for $r_{12(13)} \rightarrow \infty$ and $r_{13(23)}=0$.
%In this situation one of the IXs escapes to infinity, thus making the bound state of three IXs impossible.
\label{fig:3bodies}}
\end{figure}

Next, we consider the possibility of formation of  bound complexes that contain more than two IXs.
Consider the situation depicted in Fig.~\ref{fig:3bodies}(a), where one IX resides in the upper bilayer at a lateral location $\bm{r}_1$, and two other IXs reside in the lower bilayer at lateral locations $\bm{r}_2$ and $\bm{r}_3$ respectively.
Denoting $r_{ij} = | \bm{r}_i - \bm{r}_j|$, we can describe the total electrostatic interaction energy by
\begin{align}
	\nonumber	U_{3}(r_{12},r_{13},r_{23}) & = U_d(r_{12}) + U_d(r_{13}) \\
    & + \frac{2e^{2}}{\eps}\left(\frac{1}{r_{23}} - \frac{1}{\sqrt{r_{23}^{2} + d_{2}^{2}}}\right).
\end{align}
The first two terms are the interaction between the two IXs at different bilayers, Eq.~\eqref{eq:1},
while the third term is the same-bilayer dipole repulsion.
This last term is minimized for fixed $r_{12}$ and $r_{13}$ when  all three excitons are arranged along a single straight line when the system is viewed from above, i.e.\ when
$\bm{r}_1 - \bm{r}_2 \parallel\bm{r}_1 - \bm{r}_3$.
In this case we have $r_{23}=r_{12}+r_{13}$.
In particular, for a symmetric configuration $r_{12}=r_{13}=r_{23}/2=r$, the 3-body potential $U_{3}(r,r,2r)$ reaches a minimum $U_{3min}<0$ at some value of $r=r_{m}$.
This might indicate that a bound complex can exist. 
%However, in Fig.~\ref{fig:3bodies} we plot the potential $U_3(r_{12},r_{13},r_{12}+r_{13})$ as a function of the two lateral distances $r_{12}$ and $r_{13}$.
However, as can be seen in Fig.~\ref{fig:3bodies}, $U_{3min}<0$ is not a global potential minimum but rather a saddle point, and the true minimum is achieved when one of the IXs escapes to infinity. Therefore, a complex of three IXs is impossible. On the contrary, stacking more than two bilayer structures vertically, may result in more complex bound states of more than two IXs.
Since, the same argument applies to complexes with more than three IXs, we conclude that in double bilayer structures bound states of more than two excitons do not exist. Therefore only unbound IXs and IX$_M$ should be considered here.

\begin{figure}
\includegraphics[width=0.4\textwidth]{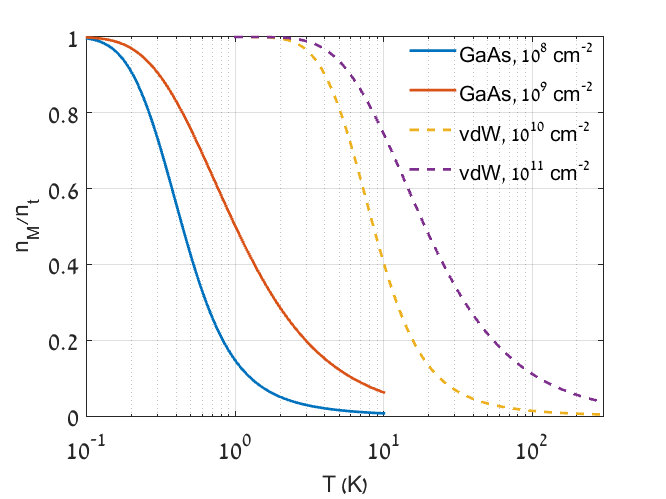}
\caption{The relative fraction of IX$_M$ density as a function of temperature for different total IX densities $n_t = n_1+n_M = n_2+n_M$, in the specific case where $n_1=n_2$.
Solid lines: a GaAs-based system with $E=0.1meV$. Dashed lines: a vdW structure with $E=2meV$.
\label{fig:saha}
}
\end{figure}

Finally, we estimate the relative fraction of IX$_M$ under realistic quasi-equilibrium conditions, in the dilute exciton gas limit where interactions are negligible and a Boltzmann distribution applies. Under these assumptions it is possible to derive a Saha-like equation (see appendix):
\begin{equation}
\frac{n_M}{n_1 n_2} = \frac{2\pi\hbar^{2}}{\mu k_B T} e^{E/k_B T}
\label{eq:IXM_saha}
\end{equation}
where $n_1, n_2$ are the unbound excitons densities in the two separate bilayers respectively, $n_M$ is the molecules density and $E$ is their binding energy (in absolute value). In Fig. \ref{fig:saha} we plot the fraction of molecules out of the total exciton density as a function of temperature for the case $n_1=n_2$. There is a significant IX$_M$ population at temperatures and densities readily achievable experimentally in both material systems. Therefore, a clear experimental signature of molecules formation is expected in the form of a new emission line in the photoluminescence spectrum: the recombination of one exciton in a molecule will result in a red shifted photoluminescence energy relative to unbound IXs due to the finite binding energy of the molecule. Owing to the residual kinetic energy of the remaining exciton, this red shift might be larger than just the molecular binding energy. As the calculated values of $E$ can be larger than the IX linewidth, this emission should be easily detectable. 
%This suggested spectroscopic signature sets the formation of the molecules apart from the more common blue shift observed in a single dipolar layer.

In conclusion, 
%we studied the formation and stability of IX$_M$ molecules in vertically coupled bilayer quantum wells.
We found that vertically coupled dipolar molecules can form in vertically coupled bilayer QWs, with a typical binding energy of few Kelvins for GaAs-based systems and up to tens of Kelvins in vdW based systems, and that such molecular species should have a significant population within a realistic range of experimental parameters, allowing a clear spectroscopic signature.
Such systems can open up opportunities for establishing quantum fluids with more complex interactions, as well as means for realizing dipolar drag and dipolar chains between two or more stacked layers of dipoles. As the observation of single IX in a trap was recently reported \cite{schinner_confinement_2013} in a single GaAs bilayer structure, observing single trapped IX$_M$ molecules in a similar trap geometries but with a vertically coupled bilayer structure could be the next step to realize complex exciton manipulations and study dipolar molecular dynamics in a well controlled system.  

\begin{acknowledgments}
We would like to acknowledge financial support from the German DFG (grant No. SA-598/9), from the German- Israeli Foundation (GIF Grant No. I-1277-303.10/2014), and from the Israeli Science Foundation (grants No. 1319/12, 1287/15)). MK would like to thank I.\frenchspacing Klich for helpful discussions.
\end{acknowledgments}

\bibliographystyle{apsrev4-1}
\bibliography{Zotero}

\input{supp.tex}

\end{document}

%% file: supp.tex
% \documentclass[preprint, aps, prl]{revtex4-1}
% %\documentclass[11pt]{article}

% \usepackage{graphicx}
% \usepackage{color}
% \usepackage{amsmath}
% \usepackage{fullpage}

% %\linespread{1.3}

% \begin{document}

% %\title{Supplementary Information}
% %\author{Kobi Cohen, Yehiel Shilo, Ran Finkelstein, Ken West, Loren Pfeiffer, and Ronen Rapaport}
% %{\large \today}
% %\maketitle

%%%%%%%%%% Merge with supplemental materials %%%%%%%%%%
\pagebreak
\widetext

\begin{center}
\textbf{\large Appendix: equilibrium density of IX$_{M}$}
\end{center}
%%%%%%%%%% Merge with supplemental materials %%%%%%%%%%

%%%%%%%%%% Prefix a "A" to all equations, figures, tables and reset the counter %%%%%%%%%%
\setcounter{equation}{0}
\setcounter{figure}{0}
\setcounter{table}{0}
\makeatletter
\renewcommand{\theequation}{A\arabic{equation}}
\renewcommand{\thefigure}{A\arabic{figure}}
% \renewcommand{\bibnumfmt}[1]{[A#1]}
% \renewcommand{\citenumfont}[1]{A#1}
%%%%%%%%%% Prefix a "A" to all equations, figures, tables and reset the counter %%%%%%%%%%

%\section{X - X$_{2}$ equilibrium}

In the case of low density both exciton and exciton molecule gases are non-degenerate, and multi-particle interactions are small \cite{laikhtman_exciton_2009}.
The free energy of exciton gases in the first and second bilayers are
\begin{subequations}
\begin{eqnarray}
&& F_{1}(n_{1},T) = - An_{1}T \ln\frac{2emT}{\pi\hbar^{2}n_{1}} \ , 
\label{eq:xx2e.1a} \\
&& F_{2}(n_{2},T) = - An_{2}T \ln\frac{2emT}{\pi\hbar^{2}n_{2}} \ ,
\label{eq:xx2e.1b}
\end{eqnarray}
\label{eq:xx2e.1}
\end{subequations}
where $A$ is the cloud area and $n_{1}, n_{2}$ are the exciton densities in the different layers respectively (here $e$ is the natural logarithm base and a unit system with $k_B=1$ is utilized). Compared to a single-component gas, a factor of 4 appears in the argument of the logarithm due to 4 possible spin states of excitons.

Similarly, it is possible to write the free energy of exciton molecular gas with density $n_{M}$:
\begin{equation}
F_{M}(n_{M},T) = - An_{M}E - An_{M}T \ln\frac{16emT}{\pi\hbar^{2}n_{M}} \ ,
\label{eq:xx2e.2}
\end{equation}
where $E$ is the molecular binding energy. Note that there are 16 possible spin state of exciton molecules and that the mass of a molecule is $2m$.

Excitons in both layers can be pumped separately and their densities are independent. The total density of excitons in each layer is then
\begin{equation}
n_{1t} = n_{1} + n_{M} \ , \hspace{1cm} n_{2t} = n_{2} + n_{M} \ .
\label{eq:xx2e.3}
\end{equation}
Given $n_{1t}$ and $n_{2t}$, the condition of equilibrium is a minimum of the total free energy,
\begin{equation}
F = F_{1} + F_{2} + F_{M} \ ,
\label{eq:xx2e.4}
\end{equation}
that is the chemical potential of the molecular gas equals the sum of chemical potentials of the exciton gases:
\begin{equation}
\mu_{1} + \mu_{2} = \mu_{M} \ ,
\label{eq:xx2e.5}
\end{equation}
where
\begin{subequations}
\begin{eqnarray}
&& \mu_{1} = \frac{1}{A} \ \frac{\partial F_{1}}{\partial n_{1}} = - T \ln\frac{2mT}{\pi\hbar^{2}n_{1}} \ ,
\label{eq:xx2e.6a} \\
&& \mu_{2} = \frac{1}{A} \ \frac{\partial F_{2}}{\partial n_{2}} = - T \ln\frac{2mT}{\pi\hbar^{2}n_{2}} \ ,
\label{eq:xx2e.6b} \\
&& \mu_{M} = \frac{1}{A} \ \frac{\partial F_{M}}{\partial n_{M}} = - E - T \ln\frac{16mT}{\pi\hbar^{2}n_{M}} \ .
\label{eq:xx2e.6c}
\end{eqnarray}
\label{eq:xx2e.6}
\end{subequations}
Substitution of Eq.(\ref{eq:xx2e.6}) in Eq.(\ref{eq:xx2e.5}) gives
\begin{equation}
n_{XM} = \frac{4\pi\hbar^{2}}{mT} \ e^{E/T} n_{X1}n_{X2} \ .
\label{eq:xx2e.7}
\end{equation}
Substituting the reduced mass $\mu = m/2$ yields Eq.(4) of the main text. %\eqref{eq:IXM_saha}
Elimination of $n_{1}$ and $n_{2}$ with the help of Eq.(\ref{eq:xx2e.3}) leads to an equation for $n_{M}$:
\begin{equation}
n_{M}^{2} - (n_{1t} + n_{2t})n_{M} - \frac{mT}{4\pi\hbar^{2}} \ e^{-E/T}n_{M} + n_{1t}n_{2t} = 0 \ .
\label{eq:xx2e.8}
\end{equation}

The solution to Eq.(\ref{eq:xx2e.8}) is
\begin{eqnarray}
n_{M} & = &
    \frac{n_{1t} + n_{2t}}{2} + \frac{mT}{8\pi\hbar^{2}} \ e^{-E/T} -
    \sqrt{\left(\frac{n_{1t} + n_{2t}}{2} + \frac{mT}{8\pi\hbar^{2}} \ e^{-E/T}\right)^{2} - n_{1t}n_{2t}}
% \nonumber \\ & = &
%     \frac{n_{X1t} + n_{X2t}}{2} + \frac{mT}{8\pi\hbar^{2}} \ e^{-E/T} 
% \nonumber \\ && \hspace{1cm} -
%     \sqrt{\frac{(n_{X1t} - n_{X2t})^{2}}{4} + \frac{mT}{8\pi\hbar^{2}} \ (n_{X1t} + n_{X2t})e^{-E/T} +
%     \left(\frac{mT}{8\pi\hbar^{2}}\right)^{2} e^{-2E/T}} \ ,
\label{eq:xx2e.9}
\end{eqnarray}
where the sign is chosen to satisfy the condition that at $T=0$ the maximun possible number of molecules is formed, that is $n_{M}=\min(n_{1t},n_{2t})$.
%Lower bound for $n_{XM}$
%\begin{eqnarray}
% n_{XM} & = & 
%     \frac{n_{X1t}n_{X2t}}
%     {\displaystyle 
%     \frac{n_{X1t} + n_{X2t}}{2} + \frac{mT}{8\pi\hbar^{2}} \ e^{-E/T} +
%     \sqrt{\left(\frac{n_{X1t} + n_{X2t}}{2} + \frac{mT}{8\pi\hbar^{2}} \ e^{-E/T}\right)^{2} - n_{X1t}n_{X2t}}}
% \nonumber \\ & > &    
%     \frac{n_{X1t}n_{X2t}}{\displaystyle n_{X1t} + n_{X2t} + \frac{mT}{4\pi\hbar^{2}} \ e^{-E/T}} \ .
% \label{eq:xx2e.10}
% \end{eqnarray}

The above result is valid as long as the exciton gases are non-degenerate, that is,
\begin{equation}
n_{1t}, n_{2t} \ll \frac{mT}{2\pi\hbar^{2}} \ .
\label{eq:xx2e.11}
\end{equation}
In GaAs DQWs with an exciton mass $m=0.21m_0$, temperature $T=1$K, and $E=0.1$meV we have:
\begin{equation}
\frac{mT}{2\pi\hbar^{2}} = 3.8 \times 10^{9} \text{cm}^{-2} \ , \hspace{1cm} e^{-E/T} = 0.31 \ , \hspace{1cm}
    \frac{mT}{2\pi\hbar^{2}} \ e^{-E/T} \approx 1.2 \times 10^{9} \text{cm}^{-2} \ .
\label{eq:xx2e.12}
\end{equation}
For example, exciton concentrations of $n_{1t}=n_{2t}=10^{9}$cm$^{-2}$ meet all of the above conditions and give exciton molecule concentration of  $n_{M}\approx0.5\times10^{9}$cm$^{-2}$.

% \bibliographystyle{apsrev4-1}
% %\bibliographystyle{unsrt}
% \bibliography{Zotero}